\documentclass[a4paper,11pt]{article}
\usepackage{pos}
\usepackage{multirow}
\usepackage{amsmath}
\usepackage{booktabs}

\def\CC{{C\nolinebreak[4]\hspace{-.05em}\raisebox{.4ex}{\tiny\bf ++}}}

\title{
Static compilation of Julia packages for integration with existing HEP codebases: a case study with JetReconstruction.jl }
 \ShortTitle{Static compilation of Julia packages}

\author[a]{Mateusz Jakub Fila}
\author[b]{Graeme A Stewart}

\affiliation[a]{CERN, Switzerland}

\affiliation[b]{DESY, Germany}

\emailAdd{mateusz.jakub.fila@cern.ch}

\abstract{The Julia programming language is considered a strong contender as a future language for high-energy physics (HEP) computing. However, transitioning to the Julia ecosystem will be a long process and interoperability between Julia and \CC\ is required. So far several successful attempts have been made to wrap HEP \CC\ packages for use in Julia. It is also important to explore the reverse direction, allowing Julia code to be called from existing HEP codebases, written primarily in \CC\ and Python, which would significantly improve potential adoption of Julia code. With recent developments in Julia enabling it to produce statically compiled code, this approach is becoming increasingly feasible, and investigating this potential for the benefit of HEP community is the focus of this work.

This work presents a case study of statically compiling the JetReconstruction.jl package - a highly performant, native Julia implementation of sequential jet reconstruction algorithms. Two different backends for Julia code compilation are compared: the existing PackageCompiler.jl and the new static compilation feature of Julia language, which is one of the major improvements in the Julia v1.12 release. The performance of the statically compiled JetReconstruction.jl is compared with both native Julia code and \CC\ FastJet.}

\begin{document}
\maketitle

\section{Introduction}

The Julia programming language \cite{Julia} has attracted significant interest in scientific computing due to its combination of high-level language features and performance approaching that of \CC. In high-energy physics (HEP), several studies have explored Julia as a potential future language for data analysis, simulation and event reconstruction software \cite{Eschle2023,JuliaInHEP}.

JetReconstruction.jl \cite{JetReconstruction} is a native Julia implementation of sequential jet clustering algorithms inspired by FastJet \cite{FastJet}. It demonstrates that idiomatic Julia code, with minimal low-level optimization, can reach or exceed the single-threaded performance of established \CC\ implementations while allowing developers to write simpler and more expressive code.

Integration of Julia-based algorithms into existing HEP software stacks remains challenging. Most existing HEP software stacks are \CC-based, and adopting Julia requires either full migration or robust interoperability mechanisms. While wrapping \CC\ libraries for use in Julia is now well established \cite{WrapCxx}, the other direction, calling Julia packages from \CC, is less explored and requires careful handling of runtime dependencies. In addition, Julia is primarily a just-in-time (JIT) compiled language, where method specialization and native code generation occur at runtime. Although this design enables high performance in long-running processes, it introduces non-negligible startup and first-execution compilation overhead. In high-energy physics computing environments, particularly distributed grid or batch systems, overhead can be repeatedly incurred and may hinder efficient resource utilization.

Recent developments in Julia v1.12 \cite{juliac} introduced static (ahead-of-time, AOT) compilation capabilities, enabling Julia functions to be compiled into standalone shared libraries or executables. This opens a potential pathway for integrating Julia-based algorithms into existing \CC\ codebases.
This work investigates this approach using JetReconstruction.jl as a case study, focusing on whether statically compiled Julia code could be realistically integrated and used in HEP environments with constraints on latency, determinism, and binary size.

\section{C-interface}

Ahead-of-time compilation of Julia code requires the external interface to be expressed through functions annotated with Julia's \texttt{@ccallable} mechanism. Such functions must expose only C-compatible types in their signatures, including primitive numeric types, pointers and simple structures. Direct interoperability with \CC\ types is not supported and therefore any interaction between a \CC\ application and Julia code must pass through a C-compatible interface.

For JetReconstruction.jl, this requirement is particularly relevant because the public API consists of multiple operations acting on the outputs of previous reconstruction stages rather than a single monolithic function call. As a result, the design of the interface must carefully consider how data is represented and transferred across the Julia--C boundary.

One challenge is that the memory layout of Julia data structures is not always suitable for direct use from C. While simple immutable structures can be mapped to equivalent C representations, the memory layout of complex data structures is an implementation detail of Julia, making them unsuitable for direct external access and difficult to exchange efficiently across the language boundary.

A second challenge arises from Julia's garbage-collected memory model. Dynamically allocated objects remain owned by the Julia runtime even when pointers to their contents are passed to C. Consequently, memory referenced by external code may be reclaimed by the garbage collector unless additional precautions are taken to preserve object lifetimes.

The current JetReconstruction.jl interface addresses these issues by using immutable structures with stable memory layout in the public interface and copying data from Julia-managed arrays into manually managed buffers before exposing them to C. This approach provides clear ownership semantics and avoids interactions with the garbage collector after control is returned to the caller. An alternative design would be to operate directly on buffers allocated by the \CC\ side and perform all computations in-place. However, such a model would require substantial changes to the internal structure of JetReconstruction.jl and was therefore not pursued in this work.

\section{Compilation}

Two approaches for ahead-of-time (AOT) compilation of Julia code were investigated in this work. The first is PackageCompiler.jl \cite{PackageCompiler}. The second is the new static compiler \texttt{juliac} introduced in Julia v1.12, exposed through the JuliaC.jl \cite{JuliaC.jl} package. 

Both approaches can produce shared libraries exposing a C-compatible API through functions annotated with Julia's \texttt{@ccallable} interface. However, neither currently guarantees complete elimination of runtime compilation for arbitrary Julia code.

The PackageCompiler.jl relies on tracing representative workloads to compile relevant code-paths and include them in produced binary. Any code-paths invoked at runtime but not traced during compilation will be JIT-compiled at runtime.
A major feature of the new compiler is support for code trimming. During compilation, static analysis is used to identify unreachable methods allowing large parts of the Julia runtime to be removed from the final binary considerably reducing the size. However, trimming currently supports only a restricted subset of the Julia language and remains experimental.

To investigate the feasibility of trimmed builds, an effort was made to adapt JetReconstruction.jl to satisfy the current compiler requirements. Several small modifications were introduced to improve type stability and eliminate patterns that prevented successful static analysis. Despite these changes, a fully trimmed build could not be achieved.
The first limitation is that LoopVectorization.jl \cite{LoopVectorization}, which is used by JetReconstruction.jl to optimize performance-critical loops, is not currently compatible with trimming.
The second limitation arises from exception handling. At the boundary between Julia and C, exceptions must be caught and translated into error codes or other C-compatible error reporting mechanisms. The current implementation of trimming does not support exception handling.
Currently trimming appears to be very restrictive and enable it may require intrusive changes in existing packages.

Consequently, all results presented in this work were obtained using non-trimmed builds. Nevertheless, ongoing development of the Julia compiler is expected to relax many of these restrictions in future releases.

The practical consequences of these limitations can be seen in the compilation metrics summarized in Table~\ref{tab:jetreconstruction_compile}. While both Julia compilation backends successfully produced deployable shared libraries, the resulting binaries remain substantially larger than the corresponding FastJet build and require significantly longer compilation times.

\begin{table}[!htb]
    \centering
    \caption{Compilation times and binary sizes for JetReconstruction.jl and FastJet 3.5.1. JetReconstruction.jl was compiled using JuliaC.jl and PackageCompiler.jl with Julia v.1.12.6.}

\begin{tabular}{cccc}
\toprule
\textbf{Code} & \textbf{Compiler} & \textbf{Compilation time} & \textbf{Binary size} \\
\midrule
\multirow{2}{*}{JetReconstruction.jl} 
& JuliaC.jl & $\sim90$ s & $\sim300$ MB \\
& PackageCompiler.jl & $\sim800$ s & $\sim375$ MB \\
\midrule
FastJet 3.5.1
& gcc 14.2 & $\sim20$ s & 8 MB \\
\bottomrule
\end{tabular}
\label{tab:jetreconstruction_compile}
\end{table}

\section{Runtime performance}

The single-threaded runtime performance of statically compiled JetReconstruction.jl was evaluated against both JetReconstruction.jl executed natively with Julia REPL, and FastJet.

\begin{figure}[!htb]
    \centering
    \includegraphics[width=0.9\linewidth]{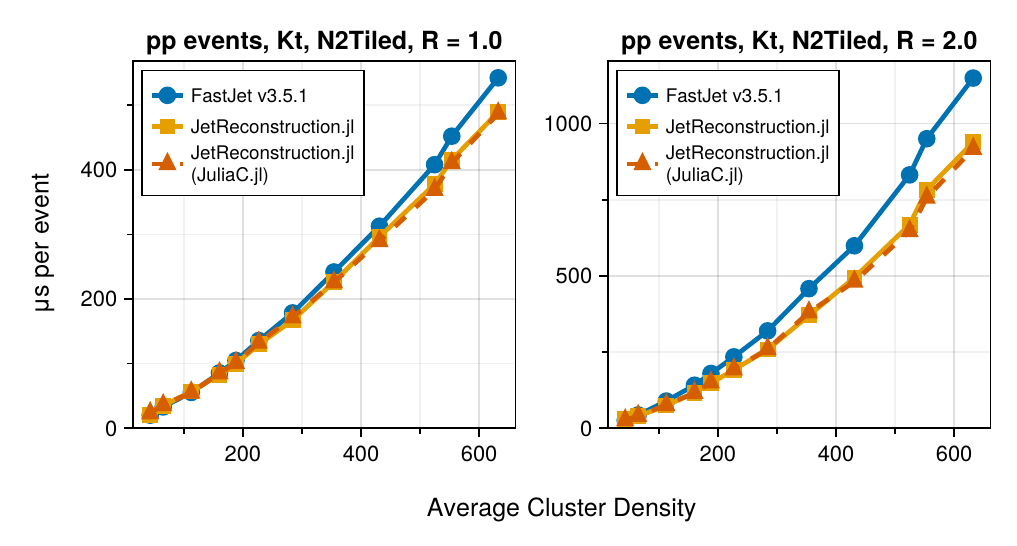}
    \caption{Comparison of jet reconstruction time for different cluster densities with FastJet 3.5.1, JetReconstruction.jl and JetReconstruction.jl C-bindings AOT-compiled with JuliaC.jl. Single-threaded $pp$ kT-algorithm and N2Tiled strategy at different jet radius values are used. JetReconstruction.jl exhibits comparable or better performance than FastJet. The overhead of JuliaC.jl compiled C-bindings is minimal with respect to native JetReconstruction.jl.}
    \label{fig:jetreconstruction_perf_c}
\end{figure}

The introduction of a C-compatible interface and AOT-compilation introduces an additional data-marshalling step between the \CC\ application and the Julia implementation. Despite this additional work, the measured overhead is small for the used workload. Figure~\ref{fig:jetreconstruction_perf_c} shows that the performance of the JuliaC.jl-compiled bindings remains very close to that of the native Julia implementation and continues to outperform FastJet.

\begin{figure}[!htb]
    \centering
    \includegraphics[width=0.55\linewidth]{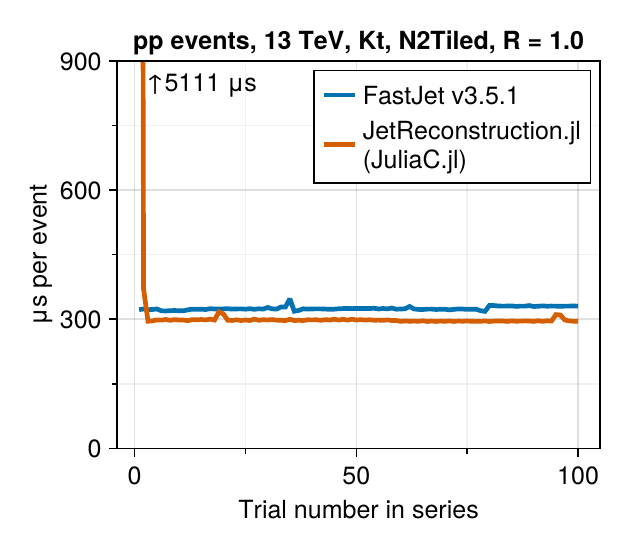}
\caption{Jet reconstruction time as a function of trial number for $pp$ kT-algorithm with N2Tiled strategy. FastJet 3.5.1 and JetReconstruction.jl C-bindings compiled with JuliaC.jl are compared. Each trial reconstructs the same sequence of 100 events within a single process. The first execution of JetReconstruction.jl C-bindings is significantly slower due to residual JIT-compilation. Subsequent executions avoid JIT-overhead and achieve lower runtime than FastJet.}
    \label{fig:jetreconstruction_trials}
\end{figure}

A more significant limitation arises from residual just-in-time (JIT) compilation. Although the code is statically compiled, the process does not yet eliminate all runtime compilation. Consequently, the first invocation of the reconstruction routines exhibits substantially higher latency than subsequent calls. This behaviour is illustrated in Figure~\ref{fig:jetreconstruction_trials}, where the first execution is noticeably slower than the remaining measurements.

PackageCompiler.jl largely avoids this issue when provided with a sufficiently complete precompilation workload, although its behaviour depends on how accurately the workload captures the code paths exercised at runtime. The JuliaC.jl-compiled version, in contrast, consistently exhibits a residual initialization penalty during the first invocation, but thereafter reaches steady-state performance comparable to the native Julia implementation.

Overall, the results indicate that AOT-compiled JetReconstruction.jl can preserve most of the performance advantages of the original package while introducing only modest overheads associated with the C-interface. However, the remaining first-call latency caused by residual JIT compilation remains a challenge for deployment in latency-sensitive environments such as trigger and online reconstruction systems.

\section{Conclusions}

This study demonstrates that Julia can be used to implement high-performance HEP algorithms that match or exceed the performance of established \CC\ implementations such as FastJet. However, integrating such code into existing \CC-based HEP frameworks via static compilation remains challenging.

Both evaluated AOT compilation approaches, PackageCompiler.jl and JuliaC.jl, are capable of producing usable shared libraries, but they do not yet provide a fully production-ready solution. In particular, issues related to binary size, compilation overhead, and incomplete elimination of runtime JIT-compilation behaviour remain unresolved.

The C-interface design requires careful upfront planning. Data structures that are convenient in idiomatic Julia do not always translate cleanly into C-compatible layouts, and retrofitting an existing package for interoperability introduces non-trivial engineering constraints.

Overall, static compilation of Julia code is a promising direction for interoperability with \CC\ frameworks, but further development of tooling, language constraints, and runtime predictability is required before such approaches can be considered production-ready for large-scale HEP systems.

\section{Notice}

The results presented at ACAT 2025 were obtained using a pre-release version of Julia 1.12.0-rc1. Since the conference, both the Julia toolchain and the static compilation ecosystem have evolved. In the present proceedings, updated results are reported using Julia 1.12.6. In addition, the juliac compiler is referenced via the JuliaC.jl package, which provides more ergonomic entry point, than pure juliac which was used in the original study. These updates do not qualitatively change the conclusions.

\section*{Acknowledgments}

This work has been partially funded by the Eric \& Wendy Schmidt Fund for Strategic
Innovation through the CERN Next Generation Triggers project under grant agreement
number SIF-2023-004.

The work has been supported by the CERN Strategic Programme on
Technologies for Future Experiments. \href{https://ep-rnd.web.cern.ch/}{https://ep-rnd.web.cern.ch/}

%
\bibliographystyle{JHEP}
\bibliography{references}

@article{JetReconstruction,
  author = {Stewart, Graeme Andrew and Ganguly, Sanmay and Ghosh, Sattwamo and Gras, Philippe and Krasnopolski, Atell},
  title = {{Fast Jet Finding in Julia}},
  DOI= "10.1051/epjconf/202533701067",
  url= "https://doi.org/10.1051/epjconf/202533701067",
  journal = {EPJ Web Conf.},
  year = 2025,
  volume = 337,
  pages = "01067",
}

@article{Julia,
author = {Bezanson, Jeff and Edelman, Alan and Karpinski, Stefan and Shah, Viral B.},
doi = {10.1137/141000671},
journal = {SIAM Review},
month = sep,
number = {1},
pages = {65--98},
title = {{Julia: A fresh approach to numerical computing}},
volume = {59},
year = {2017}
}

@article{FastJet,
    author = "Cacciari, Matteo and Salam, Gavin P. and Soyez, Gregory",
    title = "{{FastJet User Manual}}",
    eprint = "1111.6097",
    archivePrefix = "arXiv",
    primaryClass = "hep-ph",
    reportNumber = "CERN-PH-TH-2011-297",
    doi = "10.1140/epjc/s10052-012-1896-2",
    journal = "Eur. Phys. J. C",
    volume = "72",
    pages = "1896",
    year = "2012"
}

@misc{LoopVectorization,
  author       = {JuliaSIMD},
  title        = {{LoopVectorization.jl}},
  url          = {https://github.com/JuliaSIMD/LoopVectorization.jl},
  note         = {Accessed: 2025-09-26},
}

@misc{JuliaC.jl,
  author       = {JuliaLang},
  title        = {{JuliaC.jl}},
  url          = {https://github.com/JuliaLang/Juliac.jl},
  note         = {Accessed: 2026-06-01},
}

@misc{WrapCxx,
  author       = {JuliaInterop},
  title        = {{WrapCxx.jl}},
  url          = {https://github.com/JuliaInterop/CxxWrap.jl},
  note         = {Accessed: 2026-06-23},
}

@misc{juliac,
  author       = {Jeff Bezanson and Gabriel Baraldi},
  title        = {{New Ways to Compile Julia}},
  year         = {2024},
  url          = {https://www.youtube.com/watch?v=MKdobiCKSu0},
  note         = {Accessed: 2025-09-26},
 howpublished = {Talk at JuliaCon 2024},
}

@article{JuliaInHEP,
    author = "Stewart, Graeme Andrew and others",
    title = "{{Julia in HEP}}",
    eprint = "2503.08184",
    archivePrefix = "arXiv",
    primaryClass = "hep-ex",
    doi = "10.1051/epjconf/202533701266",
    journal = "EPJ Web Conf.",
    volume = "337",
    pages = "01266",
    year = "2025"
}

@Article{Eschle2023,
author={Eschle, Jonas
and G{\'a}l, Tam{\'a}s
and Giordano, Mos{\`e}
and Gras, Philippe
and Hegner, Benedikt
and Heinrich, Lukas
and Hernandez Acosta, Uwe
and Kluth, Stefan
and Ling, Jerry
and Mato, Pere
and Mikhasenko, Mikhail
and Moreno Brice{\~{n}}o, Alexander
and Pivarski, Jim
and Samaras-Tsakiris, Konstantinos
and Schulz, Oliver
and Stewart, Graeme Andrew
and Strube, Jan
and Vassilev, Vassil},
title={Potential of the Julia Programming Language for High Energy Physics Computing},
journal={Computing and Software for Big Science},
year={2023},
month={Oct},
day={05},
volume={7},
number={1},
pages={10},
issn={2510-2044},
doi={10.1007/s41781-023-00104-x},
url={https://doi.org/10.1007/s41781-023-00104-x}
}

@misc{PackageCompiler,
  author       = {JuliaLang},
  title        = {{PackageCompiler.jl}},
  url          = {https://github.com/JuliaLang/PackageCompiler.jl},
  note         = {Accessed: 2025-09-26}
}
\end{document}